# SCIENTIFIC REPORTS

**OPEN**

# Nonstoichiometric titanium dioxide nanotubes with enhanced catalytical activity under visible light

A. A. Valeeva[1,2], E. A. Kozlova[3], A. S. Vokhmintsev[2], R. V. Kamalov[2], I. B. Dorosheva[1,2], A. A. Saraev[3], I. A. Weinstein[2] & A. A. Rempel[1,2]

The catalytic activity of nanotubular titanium dioxide films formed during the oxidation of acetone to carbon dioxide under the action of visible light with a wavelength of 450 nm was found to be approximately 2 times higher compared to standard titanium dioxide (Degussa P25). The nanotubular films were grown by the anodization of titanium foil using an original technique. Diffuse reflectance spectra of the films are attributed to enhanced activity in the visible spectrum by the nonstoichiometry of titanium dioxide near the interface between the nanotubular film and the titanium foil substrate.

Nanostructural modifications of titanium dioxide are of great research interest, particularly in the development of promising functional media for renewable energy sources (i.e., solar cells and the photochemical decomposition of water)[1–4], selective organic synthesis with atomic precision for direct C–H functionalization of a variety of organic molecules[5], the manufacture of efficient photoelectric transducers and memristor memory cells[6], the development of photocatalysts for the removal of organic impurities[7–9], and other applications[10–12]. The optical gap width of stoichiometric $TiO_2$ (>3.1 eV) determines its photocatalytic activity under the action of near -UV radiation, which constitutes only a small percent of the solar spectrum[13]. To shift the spectral response to the visible region and enhance the catalytic activity, it is reasonable to decrease the optical gap width of the material, in particular, from the generation of structural vacancies in the oxygen sublattice[12,14].

The most promising possibility for generating oxygen vacancies in titanium dioxide is the growth of nanofilms from titanium foil, as has been achieved previously[6,15,16] or by sonoelectrochemical method[17]. In this case, deficiencies in oxygen can appear in the vicinity of the nonstoichiometric contact between titanium dioxide and titanium. Some indications of this phenomenon are apparent in the results of the photocatalytic activity of nanotubular films. Indeed, an enhancement of activity by 18% in comparison to Degussa P25 under the action of UV light was registered[17]. In[16], a much higher activity level of nanotubes on Ti foil compared to than on Degussa P25 was also observed under visible light.

Therefore, the aim of the present work is to consider the growth of the nanotubular titanium dioxide film with nonstoichiometry near the interface with titanium foil in detail, and to study the photocatalytic activity of such films under visible light by a reliable technique – the oxidation of acetone to $CO_2$.

## Results and Discussion

Figures 1–4 show SEM images of the surface and a side view of the grown films as a function of variation of the anodization time. A morphological analysis of the resulting films made it possible to deduce the dependences of geometric parameters, such as the effective surface and solid fraction of the nanotubular titanium dioxide layer as a function of anodization time.

According to[18], the effective surface $S_{eff}$ and roughness factor $H$ are described by equations (1) and (2)

$$S_{eff} = H \cdot S_{film}, \qquad (1)$$

[1]Institute of Solid State Chemistry, Ural Branch of the RAS, Yekaterinburg, Russia. [2]NANOTECH Centre, Ural Federal University, Yekaterinburg, Russia. [3]Boreskov Institute of Catalysis, Siberian Branch of the RAS, Novosibirsk, Russia. Correspondence and requests for materials should be addressed to A.A.V. (email: anibla_v@mail.ru)

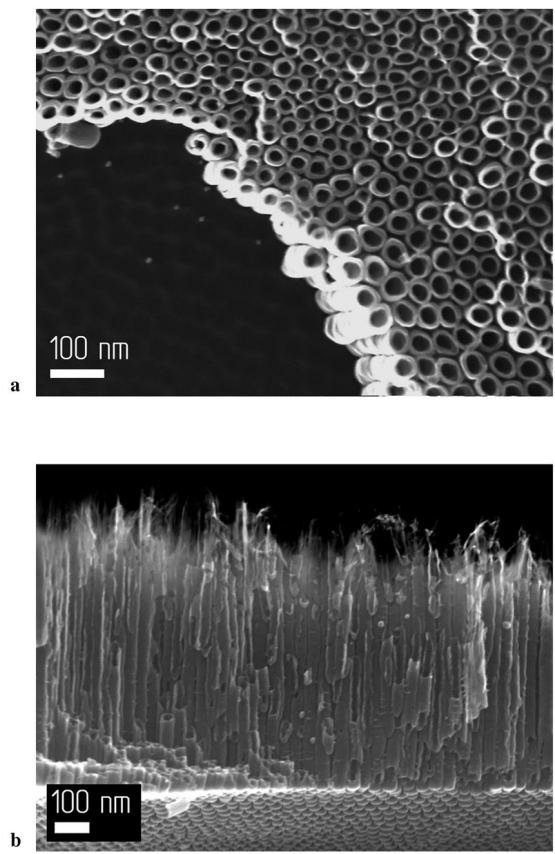

**Figure 1.** SEM images of the sample anodized for 15 min: (**a**) Surface of the sample; (**b**) side view of the sample.

$$H = \frac{8\pi LR_2}{\sqrt{3}(4R_2 - 2R_1 + y)^2},\qquad(2)$$

where $S_{film}$ is the area of anodization on which the nanotubular film was grown, $R_2 = R_1 + \frac{1}{2}w$, $R_1 = \frac{D_{in}}{2}$, $L$ is the nanotube (NT) length, $w$ is the thickness of the NT wall, $D_{in}$ is the inner diameter of the NT, and $y$ denotes the voids between nanotubes.

The solid fraction $SF$ of the nanotubular film was calculated using equation (3), as given by[18]

$$SF = \frac{\pi}{4}(D_{out}^2 - D_{in}^2)\frac{2}{\sqrt{3}(4R_2 - 2R_1 + y)^2},\qquad(3)$$

where $D_{out}$ is the outer diameter of a nanotube. According to an analysis of the morphology of the films grown at different anodization times, from 15 minutes to 6 hours, nanotube lengths range from 500 nm to 2 μm, the value of the inner diameter $D_{in}$ of nanotubes varies from 30 to 40 nm, and the outer diameter $D_{out}$ varies from 45 to 55 nm. It should be noted that at a certain time points, the growth of the oxide layer slows down considerably because the current density decreases; hence, dissolution of the oxide layer becomes more intense than oxidation. In the present study, this time point was 2 hours. After this time of anodization ($t_A$ = 120 min), an effective surface for the grown array of nanotubes reaches a maximum ($S_{eff}$ = 415 cm$^2$) at this anodization time, and a maximum length of the nanotubes of about $L$ = 2 μm was also achieved. According to equitation (3), the solid fraction $SF$ of nanotubes at an anodization time of 120 min reaches a maximum value approximately 0.41.

Therefore, an optimal anodization time to grow ordered nanotubular TiO$_2$ layers with maximum values of the $L$ and $S_{eff}$ parameters under the conditions indicated above is 120 min, which is necessary for enhancing catalytic activity.

Figure 5 depicts the XRD pattern of the nanotubular titanium dioxide film on titanium foil that was grown by anodization for 120 min. At small 2$\theta$ angles (from 18 to 33°), the intensity of the signal increased, which manifests as a diffuse halo with a maximum marked by an arrow. In addition, Fig. 5 displays the XRD pattern of the initial titanium foil on which the nanotubular titanium dioxide layer was grown by anodization. The intense diffuse halo and the absence of diffraction peaks clearly demonstrates that the nanotubular titanium dioxide layer is amorphous. This result agrees satisfactorily with independent studies[19–21].

Figure 6 shows diffuse reflectance spectra DRS of the nanotubular TiO$_2$ layer and the Degussa P25 TiO$_2$ nanopowder. In the visible spectrum at approximately 450 nm, there is a broad reduction in diffuse reflection, which is related to the nonstoichiometry of amorphous titanium dioxide. The presence of nonstoichiometry is supported by thermodynamic calculations and experimental studies using energy dispersive X-ray spectroscopy[19]. Atomic defects represented by vacancies in the amorphous network lead to the formation of energy levels in the band gap of titanium dioxide, which in its turn results in the absorption of visible light with a wavelength near 450 nm, shown by an arrow in Fig. 6. This absorption exerts a pronounced effect on the catalytic activity of the film in the visible region including the wavelength at a 450 nm, which was used in the present work for measurements of catalytic activity.

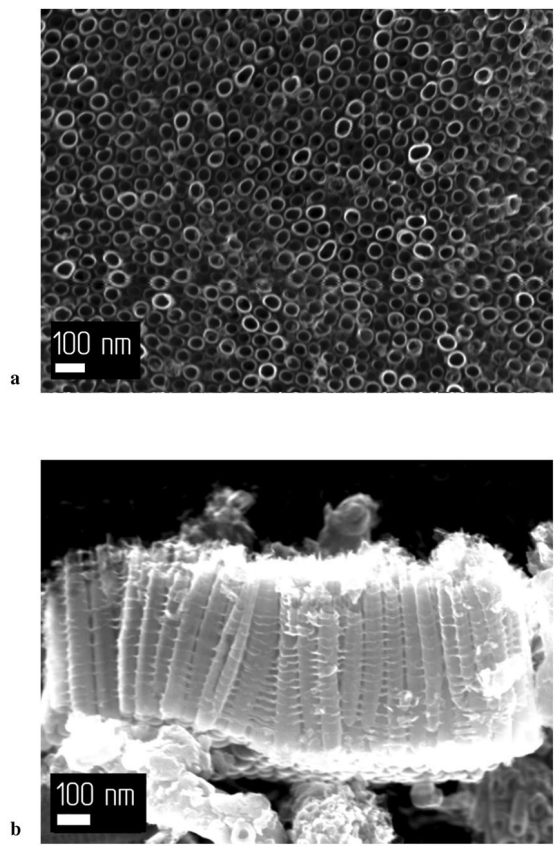

**Figure 2.** SEM images of the sample anodized for 60 min: (**a**) Surface of the sample; (**b**) Side view of the sample.

In addition, the analysis of DRS of the nanotubular $TiO_2$ layer that was anodized for 120 min allowed us to determine the optical gap width by the following technique. It was assumed that the relationship between the band gap width of a semiconductor $E_g$ and the optical absorption coefficient $\alpha$ has a Tauc function form[21–23], as given by equation (4)

$$(h\nu\alpha)^{1/n} = A(h\nu - E_g), \qquad (4)$$

where $h$ is the Planck constant, $\nu$ is the frequency of electromagnetic waves, and $A$ is a constant. In the case of nanotubular $TiO_2$ layers, the exponent $n = 2$ for indirect allowed transitions, and $n = \frac{1}{2}$ for direct ones. To find the band gap width, the diffuse reflectance spectrum was converted into a Kubelka-Munk function[24,25], as given by equation (5).

$$F(R_\infty) = \frac{(1 - R_\infty)^2}{2R_\infty} \qquad (5)$$

where $R_\infty$ is the diffuse reflection of the oxide layer relative to a white body, and it is known that $F(R_\infty)$ is proportional to the coefficient $\lambda$[25]. In this situation, equation (4) takes the form of equation (6).

$$(h\nu F(R_\infty))^{1/n} = A(h\nu - E_g) \qquad (6)$$

Figure 7 shows the Kubelka-Munk function for indirect allowed transitions and the approximation of the dependence by linear functions for a nanotubular $TiO_2$ layer (top) and the Degussa P25 $TiO_2$ nanopowder (bottom). The extrapolation with a linear function to zero absorption in the region of high photon energies shows that the maximum band gap width for a nanotubular $TiO_2$ layer anodized for 120 min is $E_g = 3.3$ eV, while $E_g = 3.0$ eV for the Degussa P25 $TiO_2$ nanopowder. An analysis of the Kubelka-Munk function in the region of low photon energies using a procedure reported in[26] revealed that the nanotubular layer contains not only a titanium dioxide phase with a wide band gap, but also the nonstoichiometric phases have an essentially narrower optical gap width, which results in the absorption of visible light, in particular with the wavelength of 450 nm (see Fig. 7, top).

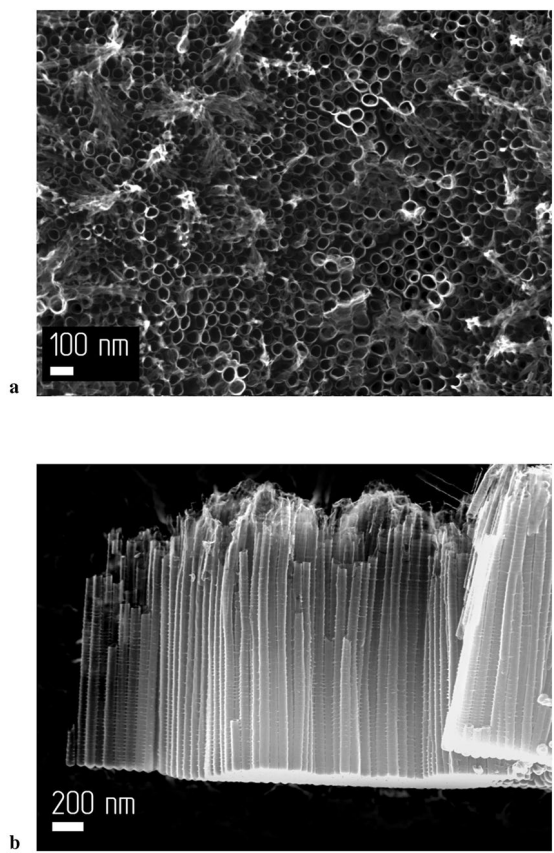

**Figure 3.** SEM images of the sample anodized for 120 min: (**a**) Surface of the sample; (**b**) Side view of the sample.

A comparison of Kubelka-Munk functions for the nanotubular amorphous film and the Degussa P25 titanium dioxide nanopowder corroborates the model of indirect transitions in the semiconductor nanotubular film. Thus, the amorphous structure of the nanotubular titanium dioxide does not produce a switch from an indirect transition (crystalline titanium dioxide) to direct one (amorphous titanium dioxide).

The Table 1 lists mean data for the catalytic activity of three different nanotubular $TiO_2$ films prepared by anodization under similar conditions for 120 min. The length of the nanotubes synthesized by this method within this time vary from 1.4 to 2 μm, and are on average 1.7 μm; however, all nanotubular layers have a similar morphology. The typical morphology is shown on Fig. 3.

The weight ($m$) of nanotubes linearly depends on their length (L) according to the formula given by equation (7)

$$m = \rho \, SF \, S_{film} L, \tag{7}$$

where $\rho$ is the density of the material. According to XRD data, the material in the present case is amorphous titanium dioxide with a density estimated at 3.6 g/cm$^3$ [27]. Therefore, the weight of nanotubes array on the area of $S_{film} = 3.8$ cm$^2$ is equal to 1.1 mg at a length of 2 μm.

The average catalytic activity measured for titanium dioxide nanotubes is approximately $4.2 \pm 0.8$ μmol/(min·g), which is more than 2 times higher compared to standard titanium dioxide (Degussa P25) (see Table 1). An analysis of the diffuse reflectance spectra of the films suggests that such a pronounced enhancement of activity are result from more efficient absorption of visible light of the films due to narrowing of the optical gap because of the nonstoichiometry of the film near the interface with the metal substrate. It should be noted that enhancement of the activity by 110% cannot be explained only by the 35% greater specific surface area of the nanotubular $TiO_2$ ($74 \pm 8$ m$^2$/g) in comparison to standard P25 $TiO_2$ ($55 \pm 5$ m$^2$/g).

A numerical comparison of the catalytic activity of different substrates (see Table 1) shows that those grown in this work are pure titanium dioxide nanotubes, which were produced more effectively than those manufactured from better-known but similar materials, and some more complicated materials. Only a few more complicated and expensive materials (e.g. 10% $UO_2(NO_3)_2/TiO_2$) are more active for oxidizing acetone, and (Au@Ag)@Au/$TiO_2$ and Au@Ag/$TiO_2$ are more active for oxidizing a 2-propanol substrate (see Table 1 and references therein).

A scheme that considers the expected band diagram of photooxidation is shown on Fig. 8. Impurity levels are electron acceptors. Energies of levels are shown on the scheme. The optical gap width of the stoichiometric $TiO_2$ is approximately 3.3 eV, and for nonstoichiometric titanium oxide ($TiO_{2-x}$) is approximately 2.8 eV. It is observed that more efficient absorption of visible light owing to an oxygen vacancy, leads to a pronounced enhancement of activity and the optical gap width decreases.

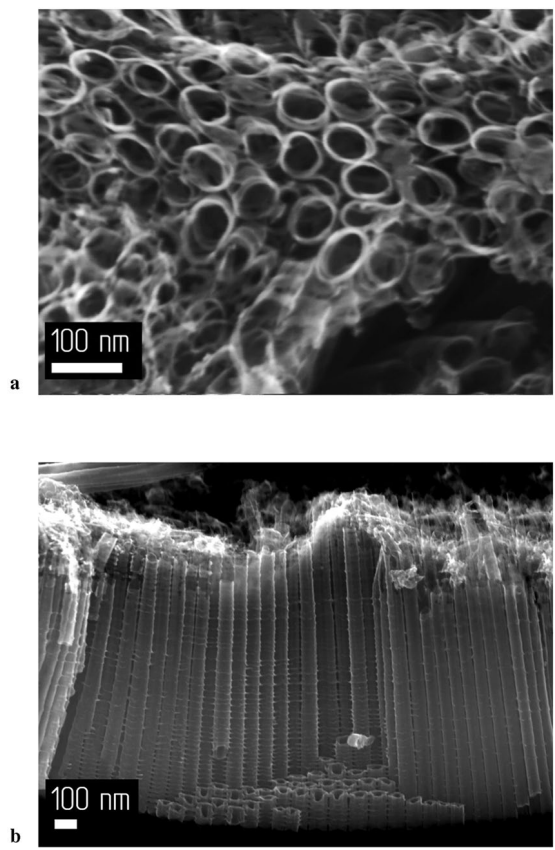

**Figure 4.** SEM images of the sample anodized for 360 min: (**a**) Surface of the sample; (**b**) Side view of the sample.

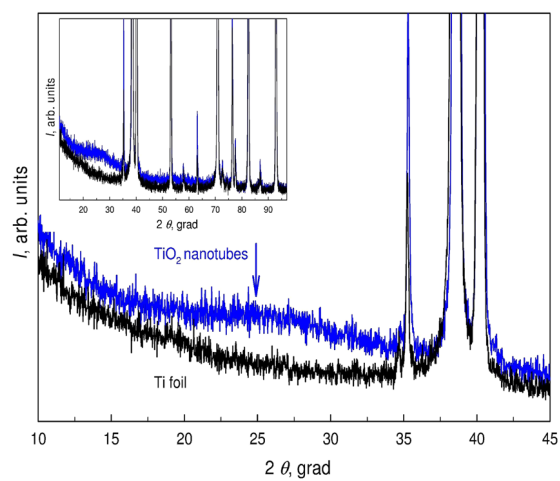

**Figure 5.** XRD pattern of the nanotubular $TiO_2$ film with the diffuse reflection maximum marked by an arrow. The XRD pattern of the initial titanium foil is displayed for comparison.

A Ti$2p$ core-level spectrum of the surface of the film is presented on Fig. 9. The spectrum is described by three doublets Ti$2p_{3/2}$ -Ti$2p_{1/2}$, corresponding to Ti in the different oxidation forms. The intense doublet Ti$2p_{3/2}$-Ti$2p_{1/2}$ with a Ti$2p_{3/2}$ peak binding energy at 459.0 eV corresponds to Ti$^{4+}$ in the structure of $TiO_2$. The second doublet Ti$2p_{3/2}$ -Ti$2p_{1/2}$ with a Ti$2p_{3/2}$ peak binding energy at 457.1 eV corresponds to Ti$^{3+}$ in the defective structure of nonstoichiometric $TiO_{2-x}$. The low intense doublet Ti$2p_{3/2}$ -Ti$2p_{1/2}$ with a Ti$2p_{3/2}$ peak binding energy at 453.6 eV corresponds to Ti° in the structure of titanium foil substrate. In the literature, the values of binding energy of Ti$2p_{3/2}$ in $TiO_2$ are given in range of 458.7–459.2 eV, while the Ti$^{3+}$ is in the range of 456.2–457.4 eV[28–30]. Therefore, XPS spectra of the films allow us to identify that the oxygen deficiency and the presence of Ti$^{3+}$ contribute to nonstoichiometry in the titanium dioxide nanotubes[31].

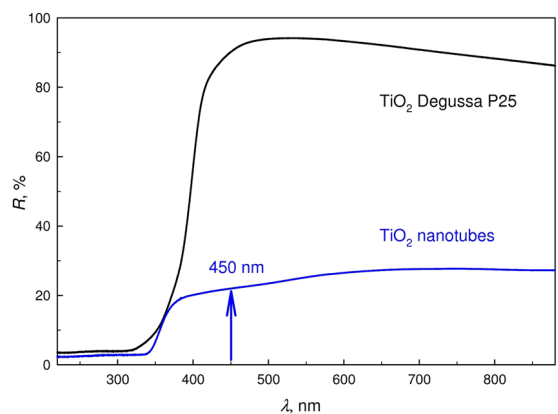

**Figure 6.** DRS of a nanotubular $TiO_2$ layer and nanostructured $TiO_2$ (Degussa P25). An arrow indicates a broad lowering of diffuse reflection in the visible spectrum near 450 nm.

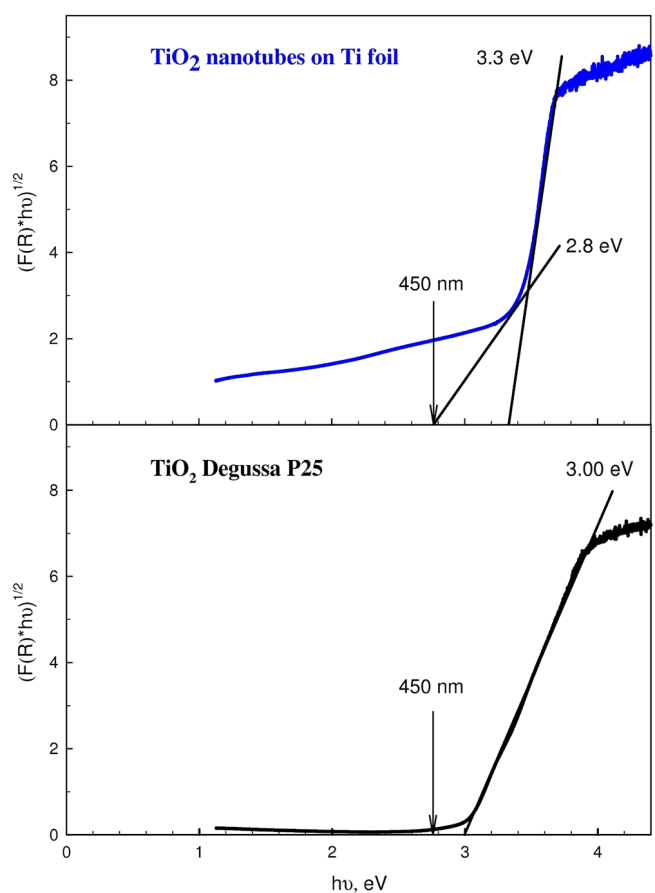

**Figure 7.** The Kubelka-Munk absorption curve for an indirect allowed transition ($n = 2$) for a nanotubular $TiO_2$ layer and Degussa P25 $TiO_2$ nanopowder. Approximations of the slope by linear functions in different regions of the absorption curve are shown. The numbers indicate energies of the optical gap width that were obtained by extrapolation to zero absorption. An arrow (450 nm) marks the energy corresponding to the wavelength of the excitation light.

## Conclusions

In the present work, the nanotubular films with a nonstoichiometric layer in the vicinity of titanium foil were grown by the anodization of titanium foil using an original technique. The highest activity of approximately $4.2 \pm 0.8$ μmol/(min·g), which is nearly twofold greater compared to standard titanium dioxide (Degussa P25), was observed for nanotubular films prepared by anodization for a period of 120 min. Such a pronounced enhancement of activity may be attributed to a more efficient absorption of visible light by the films due to narrowing of the optical gap because of difference in the nonstoichiometry of titanium dioxide near the interface between nanotubular film and the titanium foil substrate. Oxygen deficiency in the titanium dioxide nanotubes are proven by XPS measurements.

| Photocatalyst | Substrate | T, °C | Light source | Cut-off filter | W ($CO_2$)/due, $\mu$mol g$^{-1}$ min$^{-1}$ | W($CO_2$)/due, $\mu$mol min$^{-1}$ | Reference |
|---|---|---|---|---|---|---|---|
| *Organic compounds mineralization* | | | | | | | |
| **$TiO_2$ nanotubes** | acetone | 40 | 450 nm LED, 4.5 mW/cm$^2$ | | **4.2 ± 0.8** | **0.005** for 1.1 mg | **Present study** |
| Degussa P25 $TiO_2$ | | | | | 2.0 ± 0.2 | 0.018 for 9.0 mg | |
| Kronos vlp7000 | | | | | 3.9 ± 0.4 | 0.035 for 9.0 mg | |
| CQDs/$Bi_2WO_6$ | | 25 | 500 W Xe lamp, 400 mW/cm$^2$ | >420 nm | 0.4 | 0.008 | 32 |
| $Bi_2WO_6$ | | | | | 0.2 | 0.004 | |
| 10% $UO_2(NO_3)_2$/$TiO_2$ | | 40 | 1000 W Xe lamp, 17 mW/cm$^2$ | | 7.5 | 0.113 | 33 |
| 5% $UO_2(NO_3)_2$/$TiO_2$ | | | | | 3.0 | 0.045 | |
| (Au@Ag)@Au/$TiO_2$ | 2-propanol | 25 | 500 W Xe lamp, 30 mW/cm$^2$ | >440 nm | 13 | 0.002 | 34 |
| Au@Ag/$TiO_2$ | | | | | 7.0 | 0.001 | |
| BiO(ClBr)$_{0.375}$I$_{0.25}$ | | 25 | 500 W Xe lamp, 30.5 mW/cm$^2$ | 400 nm < $\lambda$ < 800 nm | 0.08 | 0.008 | 35 |
| BiO(ClBr)$_{0.5}$ | | | | | 0.04 | 0.003 | |
| *Dye degradation* | | | | | | | |
| WO$_3$@$TiO_2$-nanotubes@WO$_3$ | RhB | 25 | Xe lamp with the intensity of 250 W at 420 nm | | 0.3 | 0.017 | 36 |
| C-$TiO_2$ nanotubes | | 20 | 500 W tungsten-halogen lamp | >420 nm | 0.4 | 0.023 | 37 |
| $TiO_2$ nanotubes | | 25 | 1000 W Xe-Hg lamp | — | 0.1 | 0.0008 | 38 |
| ZnFe$_2$O$_4$-$TiO_2$ nanotubes | AOII | 30 | 500 W tungsten-halogen lamp | >400 nm | — | 0.0006 | 39 |
| GO-Ag-$TiO_2$ nanotubes | MB | 25 | 500 W tungsten-halogen lamp | | — | 0.002 | 40 |
| $TiO_2$ nanotubes | | | Ultraviolet lamp ($\lambda$ = 365 nm) | — | — | 4·10$^{-6}$ | 16 |
| $TiO_2$ nanotubes | Orange G | | Ultraviolet lamp ($\lambda$ = 254 nm) | | 2.4 | — | 17 |
| $TiO_2$ nanotubes | AO7 | 25 | Ultraviolet lamp ($\lambda$ = 325 nm) | | — | 0.0006 | 41 |
| $TiO_2$ nanotubes | MB | 20 | Ultraviolet lamp ($\lambda$ = 325 nm) | | | 0.0007 | 42 |
| GR-$TiO_2$ nanotubes | Alachlor (herbicide) | 25 | Xe lamp | — | — | 0.1 | 43 |

**Table 1.** A numerical comparison of previously published catalytic activities with data obtained from the present study.

## Methods

Nanotubular titanium dioxide films were synthesized on a Digma setup consisting of an electrochemical cell, thermostat and power supply[6]. Anodic oxidation was carried out using a substrate made of 100 $\mu$m thick titanium foil. The surface of the titanium foil was preliminarily treated with acetone and an aqueous solution of acids mixed in a volumetric ratio of $HNO_3$:HF:$H_2O$ = 6:1:18, and was then washed with distilled water. Primary anodization was performed in a 1% solution of hydrofluoric acid in ethylene glycol for 10 min at a temperature of 20 °C. This anodization was followed by etching of the titanium substrate in a solution of acids ($HNO_3$:HF:$H_2O$ = 1:1:20) for 3 min, washing in distilled water and acetone, and finally drying at room temperature. Secondary anodization was carried out potentiostatically at a voltage of 20 V and anodization times of $t_A$ = 15, 30, 60, 120, 180 or 360 min. A solution of ethylene glycol and ammonium fluoride with a concentration equal to 1 wt. % was employed as the electrolyte. All the chemical reagents used in the study were of analytical grade. The area of primary and secondary anodization for every sample was approximately 3.8 cm$^2$.

XRD analysis of the films and substrate was performed using Cu$K\alpha_{1,2}$ radiation on a Shimadzu XRD-7000 (Japan) diffractometer with the Bragg-Brentano recording geometry. XRD patterns were measured using the step-scan mode at $\Delta(2\theta)$ = 0.02° in the 2$\theta$ angular range from 10 to 100° with a long exposure time at each step. The surface and side views of the synthesized samples were examined on a SIGMA VP (Carl Zeiss) scanning electron microscope (SEM) under high vacuum using an InLens detector. The obtained SEM images were processed with Clinker C7 software (SIAMS).

DRS measurements of the films and standard were recorded on a Lambda 35 (Perkin Elmer, USA) UV-VIS spectrophotometer with a RSA-PE-20 (Labsphere) diffuse reflectance attachment.

The XPS measurements were performed on a photoelectron spectrometer (SPECS Surface Nano Analysis GmbH, Germany) equipped with a PHOIBOS-150 hemispherical electron energy analyzer, a FOCUS-500 X-ray monochromator, and a XR-50 M X-ray source with a double Al/Ag anode. The core-level spectra were obtained using monochromatic Al K$\alpha$ radiation ($h\nu$ = 1486.74 eV) and a fixed analyser pass energy of 20 eV under ultra-high-vacuum conditions. The charge correction was performed by setting the Ti$2p_{3/2}$ peak at 459.0 eV, corresponding to titanium in $TiO_2$. For detailed analysis, the spectra were fitted into several peaks after background subtraction using the Shirley method. The fitting procedure was performed using CasaXPS software. The line shapes were approximated by the product of Gaussian and Lorentz functions.

Photocatalytic activity of the films and the standard was measured in a flow reactor, and reactants were identified using an FTIR spectrometer. Standards – (titanium dioxide Degussa P25 and Kronos vlp7000) were uniformly deposited on a square glass slides ($S = 9$ cm$^2$) from aqueous suspensions, so that the surface density was 1 mg/cm$^2$. The study was performed using a light-emitting diode with a maximum wavelength of 450 nm. The other reaction condition were as follows: $T = 40$ °C, relative humidity of 20%, an acetone concentration of 800 ppm, and a flow rate of 60 mL/min. Oxidation was performed using the oxygen in air.

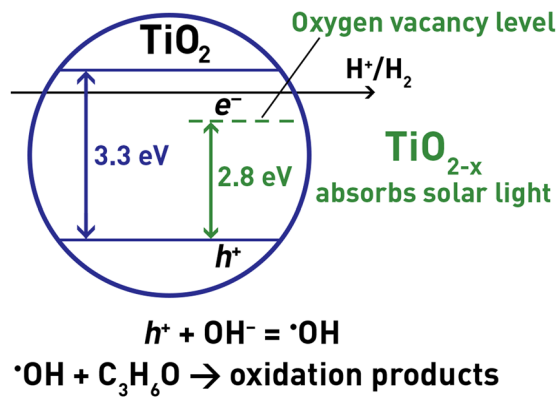

**Figure 8.** A scheme considering the expected band diagram of photooxidation.

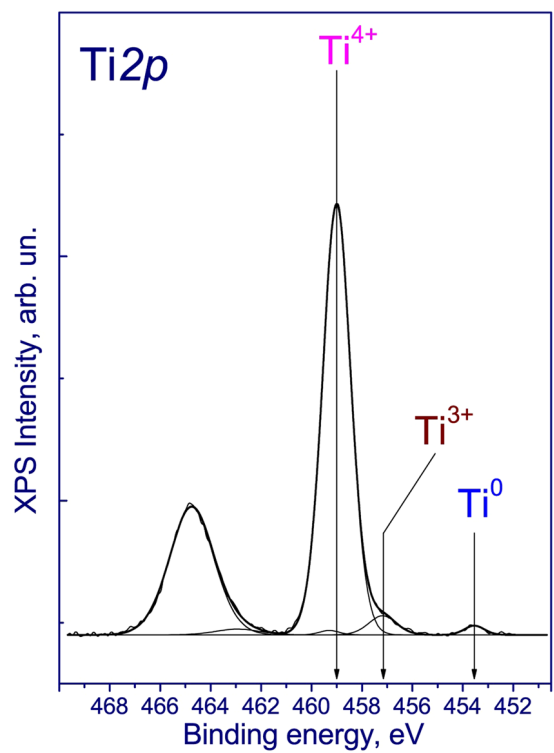

**Figure 9.** Ti$2p$ core-level spectra of the surface of the film. There are Ti$^{3+}$ ions with Ti$2p_{3/2}$ binding energy at 457.1 eV, in addition to Ti$^{4+}$ ($E_{bin} = 459.0$ eV).

The concentration of acetone was found by integrating the area of the absorption band at 1160–1265 cm$^{-1}$, and the concentration of carbon dioxide was found by integrating the area of the absorption band at 2200–2450 cm$^{-1}$. The rate of photocatalytic oxidation was calculated from the accumulation of carbon dioxide, which was a sole product of the reaction:

$$C_3H_6O + 4\,O_2 \rightarrow 3\,CO_2 + 3\,H_2O.$$

As CO$_2$ is the sole product of acetone photooxidation, the rate of acetone decomposition was one third of the rate of carbon dioxide formation rate.

### Acknowledgements
This work was partially financially supported by the Russian Foundation for Basic Research (project No. 17-03-00702) and by Act 211 Government of the Russian Federation, contract No. 02.A03.21.0006. A.S.V. and I.A.W. thank Minobrnauki initiative research project № 16.5186.2017/8.9 for support.

### Author Contributions
A.A.R. and I.A.W. designed the research. A.A.V., I.B.D., R.V.K. prepared the samples. A.A.V., E.A.K., A.S.V., A.A.S. performed the experiments. All authors wrote, read and approved the final manuscript.

### Additional Information
**Competing Interests:** The authors declare no competing interests.

**Publisher's note:** Springer Nature remains neutral with regard to jurisdictional claims in published maps and institutional affiliations.